\documentclass[11pt,twoside,letterpaper]{article} 
\usepackage{times,fancyhdr}
\usepackage{graphicx}
\makeatletter 
\def\cleardoublepage{\clearpage\if@twoside \ifodd\c@page\else%
    \hbox{}%
    \thispagestyle{empty}%
    \newpage%
    \if@twocolumn\hbox{}\newpage\fi\fi\fi} 
\makeatother 
\def\figurename{Figure}
\makeatletter
\renewcommand{\fnum@figure}[1]{\figurename~\thefigure.}
\makeatother
\def\tablename{Table}
\makeatletter
\renewcommand{\fnum@table}[1]{\tablename~\thetable.}
\makeatother
\begin{document}
\title{
{\begin{flushleft}
\vskip 0.45in
{\normalsize\bfseries\textit{Chapter~1}}
\end{flushleft}
\vskip 0.45in
%
%
%
%
\bfseries\scshape Neutron Star Crust and Molecular Dynamics Simulation}}
\author{\bfseries\itshape C. J. Horowitz$^1$\thanks{E-mail horowit@indiana.edu},\,  J. Hughto$^1$,\,  A. Schneider$^1$, D. K. Berry$^2$\\
1) Center for Exploration of Energy and Matter and Physics Department,\\ Indiana University, Bloomington, IN 47405, USA\\
2) University Information Technology Services,\\ Indiana University, Bloomington, IN 47408, USA}
\date{}
\maketitle
\thispagestyle{empty}
\setcounter{page}{1}
\thispagestyle{fancy}
\fancyhead{}
\fancyhead[L]{In: Neutron Star Crust \\ 
Editors: C.A. Bertulani and J. Piekarewicz, pp. {\thepage-\pageref{lastpage-01}}} 
\fancyhead[R]{ISBN 0000000000  \\
\copyright~2012 Nova Science Publishers, Inc.}
\fancyfoot{}
\renewcommand{\headrulewidth}{0pt}
\vspace{2in}
\vspace{.08in} \noindent \textbf{Keywords: neutron star crust, molecular dynamics, nuclear pasta, breaking strain}
%
\pagestyle{fancy}
\fancyhead{}
\fancyhead[EC]{C. J. Horowitz {\it et al.}}
\fancyhead[EL,OR]{\thepage}
\fancyhead[OC]{NS Crusts and MD}
\fancyfoot{}
\renewcommand\headrulewidth{0.5pt} 
%
\section{Introduction}

Stars are hot plasmas.  Therefore, the presence of a solid neutron star (NS) crust may seem unlikely.  Nevertheless compact stars, such as white dwarfs (WD) and NS, are so dense that this plasma can crystallize.  How do these stars freeze, and what are some of the extraordinary properties of this solid star stuff?  In this chapter we describe molecular dynamics (MD) simulations of plasma crystals and NS crust.  These simulations determine many crust properties that may be crucial for the electromagnetic, neutrino, and gravitational wave radiations of NS.  

We start in Sec. \ref{sec.freeze} describing how plasmas freeze in the laboratory, in the interior of white dwarf stars, and to form new neutron star crust.
In Sec. \ref{sec.formalism}  we describe a molecular dynamics formalism for accurately determining many crust properties.  In Sec. \ref{sec.results} we review MD results for chemical separation upon freezing, diffusion in Coulomb solids, breaking strain (strength of the NS crust), shear viscosity, and dynamical response.  Finally, Sec. \ref{sec.conclusions} contains a short summary and discusses some open questions and future challenges.             

\section{How do Stars Freeze?}
\label{sec.freeze}
How do compact stars freeze?  Surprisingly there has been little work on this fundamental question.  In this section we first examine crystallization in laboratory plasmas and then discuss freezing in white dwarfs and neutron stars. 

\subsection{Crystallization of Laboratory Plasmas}
In the laboratory, one can study complex (or dusty) plasma crystals.  Complex plasmas (CP) are low temperature plasmas containing charged microparticles, for a review see Fortov et al. \cite{fortov}.  Often the microparticles are micron sized spheres that acquire large electric charges and the strong coulomb interactions between microparticles can lead to crystallization.  Indeed plasma crystals were first observed in the laboratory in 1994 \cite{dusty_plasma}.   Complex plasmas typically differ from White Dwarf interiors and Neutron Star crusts in a number of ways.  First the microparticles feel additional fluctuating and friction forces because of interactions with the background gas.  Note that in stars, electron-ion interactions are small because of the large electron degeneracy.  Second, the Debye screening length $\lambda$, see Eq. \ref{v(r)} in Sec. \ref{sec.formalism}, is often smaller in the CP than in a star (when measured in units of the lattice spacing).  This changes the lattice type from body-centered-cubic (bcc) as expected in stars, to face-centered-cubic (fcc) or other types in a CP.  Finally in a CP there is an overall confining potential, and because of gravitational gradients it is often easier to study two-dimensional CP crystals.  



Three-dimensional CP crystals have been formed onboard the International Space Station under microgravity conditions.  Details of the experiment are presented in ref. \cite{ISS1}.  Microparticles were found in regions with fcc and hexagonal-close packing (hcp) order \cite{ISS2,ISS3}, in agreement with MD simulations \cite{ISS2}.   Khrapak et al. \cite{khrapak} studied freezing and melting of these CP crystals and found diffusion to be relatively fast so that the system remained in equilibrium.  Melting criteria for CP systems were presented by Klumov \cite{klumov0}.   It should be possible to record complete trajectories, position as a function of time, for each of the microparticles in a three dimensional CP experiment.  This would allow a very detailed dynamical study of the melting or freezing phase transitions.

\subsection{Crystallization in White Dwarf stars}

Observations of cooling White Dwarf (WD) stars provide important information on the ages and evolution of stellar systems \cite{cosmochron, garcia-berro, renedo, salaris1}.  The interior of a WD is a coulomb plasma of ions and a degenerate electron gas.  As the star cools this plasma crystallizes.  Note that WDs freeze from the center outward, because of the higher central densities.  While NS are expected to form a solid crust over a liquid core.  This crystallization can delay WD cooling as the latent heat is radiated away, see for example ref. \cite{salaris}.  Winget et al. recently observed effects from the latent heat of crystallization on the luminosity function of WDs in the globular cluster Ngc 6397 \cite{winget}.   Winget et al.'s observations may constrain the melting temperature of the carbon and oxygen mixtures expected in these WD cores.  In addition, Astro-seismology provides an alternative way to study crystallization in WD, see for example \cite{astroseismology}.    

How do stars freeze?  At present we have very little direct observational information to answer this fundamental question.  In the future we can observe many more WD systems and this should greatly increase the observational information on crystallization in WD.  

\subsection{Crystallization on Accreting Neutron Stars}

We next consider freezing in accreting NS.  Material falling on a NS can undergo rapid proton capture (or rp process) nucleosynthesis to produce a range of nuclei with mass numbers $A$ that could be as high as $A\approx 100$ \cite{rpash, rpash2}.  As this rp process ash is buried by further accretion, the rising electron fermi energy induces electron capture to produce a range of neutron rich nuclei from O to approximately Se \cite{rpash3}.  This material freezes, when the density reaches near $10^{10}$ g/cm$^3$.  In Sec. \ref{subsec.phasesep} we describe large scale MD simulations of how this complex rp process ash freezes \cite{phasesep}.  We find that chemical separation takes place and the liquid ocean is greatly enriched in low atomic number $Z$ elements, while the newly formed solid crust is enriched in high $Z$ elements.  Note that for an accreting star, there is typically a thin liquid ocean atop a solid crust that itself surrounds a dense liquid core. 

Furthermore, MD simulations, as discussed in Sec. \ref{sec.results} suggest that a regular crystal lattice forms even though large numbers of impurities may be present.  This regular crystal should have a high thermal conductivity.  We do not find an amorphous solid that would have a low thermal conductivity \cite{soliddiffusion}.

Recent X-ray observations of neutron stars, find that the crust cools quickly, when heating from extended periods of accretion stops \cite{crustcooling,crustcooling1,crustcooling2}.  This is consistent with MD simulations, and strongly favors a crystalline crust over an amorphous solid that would cool more slowly \cite{crustcooling3},  \cite{crustcooling4}.  


\section{Molecular Dynamics Formalism}
\label{sec.formalism}

Neutron star crust consists of a relativistic fermi gas of electrons, a crystal lattice of neutron rich ions, and, in general, a neutron gas.  For a review see \cite{crustreview}.       In  Section \ref{subsec.MDformalism} we  describe our MD simulation formalism and then discuss some computational considerations in Section \ref{subsec.computers}

\subsection{Classical Molecular Dynamics for Coulomb plasmas}
\label{subsec.MDformalism}
We model NS crust as a classical Coulomb plasma.  The potential between the $i^{th}$ and $j^{th}$ ions is assumed to be of screened Coulomb (Yukawa) form,
\begin{equation}
v_{ij}(r)=\frac{Z_iZ_j e^2}{r} {\rm e}^{-r/\lambda}.
\label{v(r)}
\end{equation}
Here the ions are separated by a distance $r$ and have charges $Z_i$ and $Z_j$.  The Fermi screening length $\lambda$, for cold relativistic electrons, is $\lambda^{-1}=2\alpha^{1/2}k_F/\pi^{1/2}$ where the electron Fermi momentum $k_F$ is $k_F=(3\pi^2n_e)^{1/3}$ and $\alpha$ is the fine structure constant.  The electron density $n_e$ is equal to the ion charge density, $n_e=\langle Z\rangle n$, where $n$ is the ion density and $\langle Z\rangle$ is the average charge.  Our simulations are classical and we have neglected the electron mass (extreme relativistic limit).   This is to be consistent with our previous work on neutron stars.  However, the electron mass is important at the lower densities in WD and this may change our results slightly \cite{pot1}.  We assume the ions are classical.  In general the large mass of the ions ensures that their thermal wavelengths are smaller than the inter-particle spacing.  However, quantum effects could play some role at high densities \cite{pot2},\cite{jones}.  For relativistic electrons, the ratio of $\lambda$ to the ion sphere radius $a$,
\begin{equation}
a=\Bigl(\frac{3}{4\pi n}\Bigr)^{1/3},
\end{equation}
depends only on the average charge $\langle Z \rangle$.  

The simulations can be characterized by a Coulomb parameter $\Gamma$ describing the ratio of a typical Coulomb energy to the thermal energy.  For a one component plasma (OCP), where all of the ions have charge $Z$, at a temperature $T$ we have,
\begin{equation}
\Gamma=\frac{Z^2e^2}{aT}\, .
\end{equation}
For a mixture of ions of different charges we define $\Gamma$ using an appropriate average of $Z_i^2/a_i$ where the typical distance between ions $a_i$ is expected to scale as $Z_i^{1/3}$ so that,
\begin{equation}
\Gamma= \frac{\langle Z^{5/3} \rangle e^2}{a_e T}\, .
\label{gammamix}
\end{equation} 
Here $\langle Z^{5/3} \rangle$ is an average over the ion charges, $T$ is the temperature, and the electron sphere radius $a_e$ is $a_e=(3/4\pi n_e)^{1/3}$ with $n_e=\langle Z\rangle n$ the electron density.  The one component system freezes near $\Gamma=175$, while mixtures of ions are expected to freeze at a somewhat higher $\Gamma$ \cite{WD_PRL,WD2}.
 
Time can be measured in units of one over the plasma frequency $\omega_p$.  Long wavelength fluctuations in the charge density can undergo oscillations at the plasma frequency.  This depends on the ion charge $Z$ and mass $M$.  For mixtures we define a hydrodynamical plasma frequency $\bar\omega_p$ from the simple averages of $Z$ and $M$,
\begin{equation}
\bar\omega_p=\Bigl[\frac{4\pi e^2\langle Z\rangle^2 n}{\langle M \rangle}\Bigr]^{1/2}.
\label{omega}
\end{equation}

\subsection{Computational Considerations for MD Simulations}
\label{subsec.computers}

The pure unscreened Coulomb interaction has an effectively infinite range. MD simulations involving pure Coulomb interactions must take this into account to  produce physically accurate results. Often this is done by assuming periodic boundary conditions and using Ewald summation to account for interactions with distant particles.  At the opposite extreme, systems with very short-range  interactions, such as the Lennard-Jones potential, can  be simulated with an MD code that uses a very small cut-off distance.   One assumes the interaction is zero for distances beyond the cutoff.  In these systems, each particle interacts with only a few tens to hundreds of nearby neighbors, even if the system as a whole involves millions of particles.  The screened Coulomb
interaction in our ion simulations falls between these extremes. We do not need
to account for interactions with very distant ions, yet because the screening
length $\lambda$ is rather large, a small cut-off radius will not model the physics
properly.   We find that a cutoff of about $8\lambda$ gives good physical results.   Note that a cutoff distance significantly smaller than $8\lambda$ will lead to errors in the shear modulus \cite{shearmodulus}.  We use a velocity Verlet integration scheme to advance the system in time \cite{verlet}.

Our computer code is a parallel Fortran 95 program that uses both MPI (Message Passing  Interface) and OpenMP.  The force calculation is carried out in a pair of nested DO loops. The outer loop is over what we call \emph{target} particles, while the inner loop is over \emph{source} particles. These are the same sets of particles of course, but for sake of discussion it is convenient to distinguish the particles that are being acted on and those doing the acting.

Most supercomputers at the time of this writing have a hybrid architecture. These machines consist of many individual compute nodes which are complete computers  within themselves.  Each node in turn consists of two or more processor chips sharing a common physical memory.  Processors chips in turn have several cores that share a common Level 2 cache and interface to main memory. Each core has its own program counter, floating point and integer registers, and floating point, integer and  logical units. Each core is therefore capable of running a \emph{thread}, or  light-weight process. This type of node architecture is similar to a symmetric multiprocessor (SMP), the kind of architecture for which OpenMP was designed.  On the other hand, MPI is designed for multi-node machines, where data can be shared among nodes only by passing messages. The combination of MPI and OpenMP in one code is often the best programming paradigm for multi-node systems where each node is an SMP.  For example, we have run our MPI+OpenMP code on a Cray XT5 computer called  Kraken, which consists of 9408 or more nodes, each of which has two six-core processors. We  typically use 64 nodes, placing an MPI process on each of node's two processors, and assigning on OpenMP thread to each processor's six cores. This makes a total of 768 threads.  Each MPI process gets a complete set of particles. Targets are  distributed among MPI processes, while sources are distributed among OpenMP threads. Each process lets all sources act on its targets.  

Our code scales remarkably well. Runs using 768 cores of Kraken obtain almost linear speedup on the 55296-ion runs, and only slightly worse on 27648-ion runs.  
Our longest simulation times, of order $2\times 10^6/\omega_p$, involved a week or two of computer time with 768 cores.  Note that the breaking strain simulations in Sec. \ref{subsec.breaking} used a modified version of the program SPaSM \cite{spasm}.

\section{MD Simulation Results}
\label{sec.results}
In this section we review MD simulation results for a number of neutron star crust properties.  We start in Sec. \ref{subsec.phasesep} with chemical separation that occurs as new crust forms on an accreting neutron star.  Next, in Sec. \ref{subsec.diffusion} we describe diffusion in Coulomb crystals that may be important for the structure of NS crust.  The breaking strain (strength) of NS crust is discussed in Sec. \ref{subsec.breaking}  This may be important for mountains on rotating NS that radiate gravitational waves.  The shear viscosity of nuclear pasta is discussed in Sec. \ref{subsec.viscosity}  This may be important for the damping of collective modes.  Finally we discuss the dynamical response of nuclear pasta in Sec. \ref{subsec.pastaosc}

\subsection{Phase separation in the crust of accreting neutron stars}
\label{subsec.phasesep}

Phase separation is important for white dwarf stars \cite{WD_PRL}.   As a star cools, and crystallization takes place, the crystal phase is enriched in oxygen while the liquid is enriched in carbon.  Phase separation is also important for neutron stars that accrete material from a companion.  This material can undergo nuclear reactions involving rapid proton capture (the rp process) to synthesize a variety of medium mass nuclei \cite{rpash}.  Further accretion increases the density of a fluid element until crystallization occurs near a density of $10^{10}$ g/cm$^3$.  However, molecular dynamics simulations show that this crystallization is accompanied by chemical separation.  The composition of the new solid crust is very different from the remaining liquid ocean.  This changes many properties of the crust and can impact many observables.
\begin{figure}[ht]
\center\includegraphics[width=5in,angle=0,clip=true] {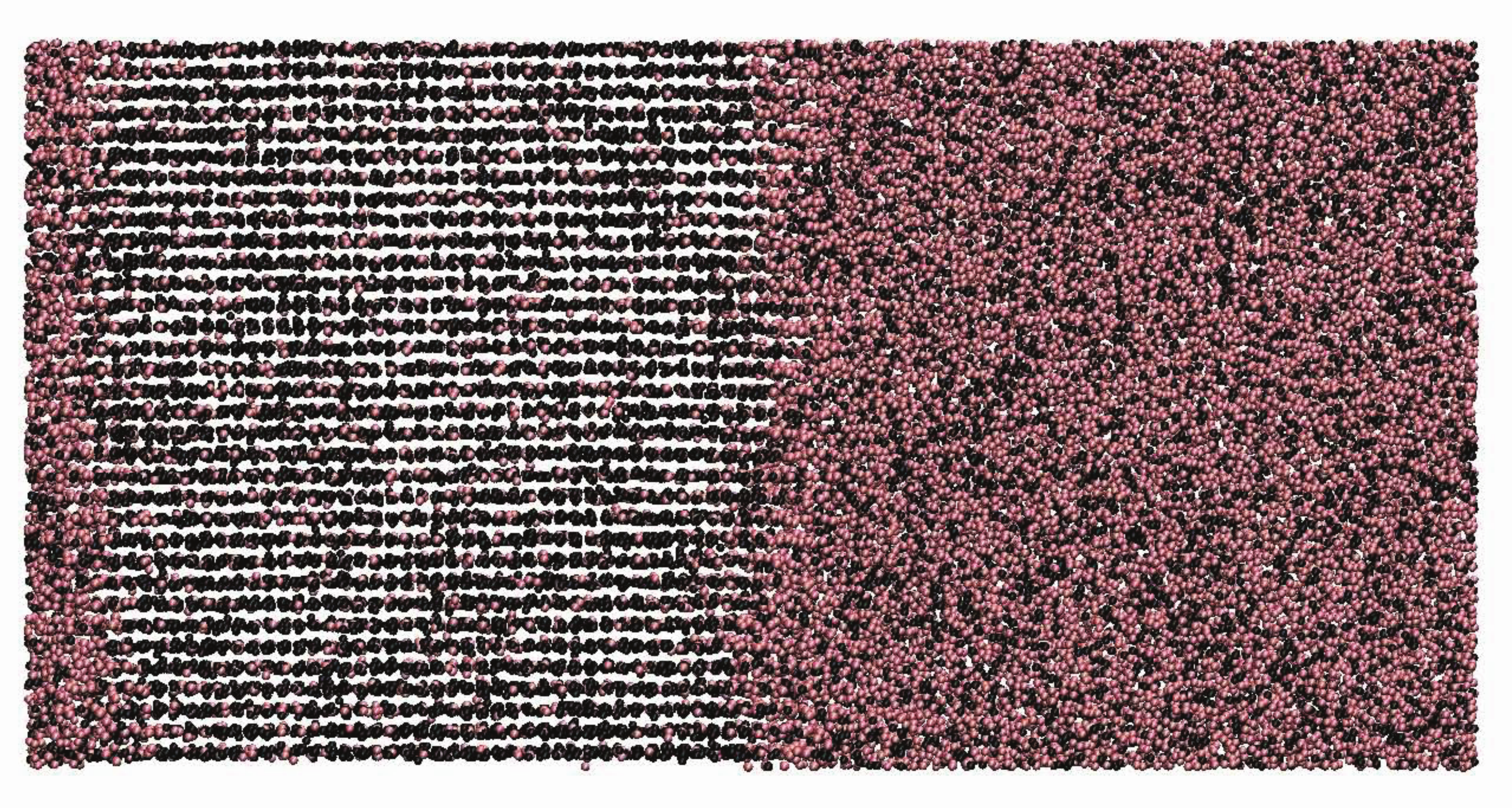}
\caption{Final configuration of carbon ions (light red) and oxygen ions (black) in a 55296 ion simulation that consisted of 75\% oxygen and 25\% carbon \cite{WD2}.  Image prepared with VMD \cite{VMD}.}
\label{Fig1}
\end{figure}


We start with the simpler two component carbon-oxygen system in the interior of a white dwarf \cite{WD_PRL,WD2}.  To determine the melting temperature and the composition of the liquid and solid phases that are in equilibrium, we perform two-phase MD simulations.  A region of solid phase is combined with a region of liquid phase, in a rectangular simulation volume, as shown in Fig. \ref{Fig1}.  Here a regular solid phase is on the left and the liquid phase is on the right.  With periodic boundary conditions, there are two liquid-solid interfaces, one near the left side and one near the center.

\begin{figure}[ht]
\begin{center}
\includegraphics[width=5.75in,angle=0,clip=true] {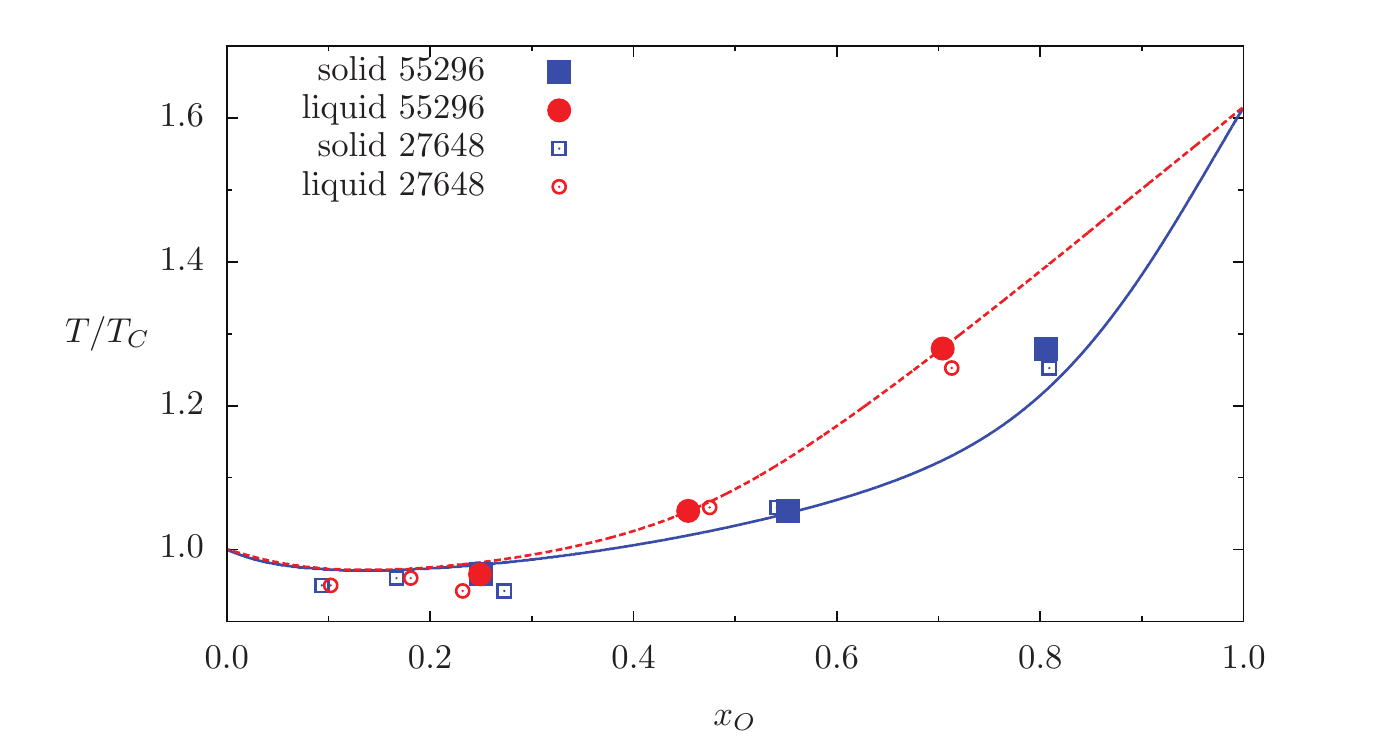}
\caption{Carbon-oxygen phase diagram plotting the composition of the liquid phase (upper red curve or circles) that is in equilibrium with the solid phase (lower blue curve or squares) at a melting temperature $T$ in units of the melting temperature for pure carbon $T_C$, from ref. \cite{WD2}.  The number fraction of oxygen ions is $x_O$.  Results from 55296 ion simulations are filled symbols while the open symbols are results with 27648 ions.  The curves are from the analytic model of Medin and Cumming \cite{medin}. }
\label{Fig2}
\end{center}
\end{figure}

 This system is then evolved for a long time while the temperature is continually adjusted to keep about half of the system solid and half liquid.  During this time carbon ions can diffuse into or out of the solid until the two phases reach chemical equilibrium.  The composition of the liquid and solid phases that are in equilibrium can be determined by simply counting the number of ions that remain in different regions of the simulation volume.  The resulting phase diagram is shown in Fig. \ref{Fig2}.  The melting temperature of an approximately 50\% carbon, 50\% oxygen mixture, that is expected in WD interiors, is seen to be only slightly higher than the melting temperature of pure carbon.  This result appears to be in good agreement with observations of WD cooling in the globular cluster NGC6397 \cite{winget}.  

\begin{figure}[ht]
\begin{center}
\includegraphics[width=3.75in,angle=0,clip=true] {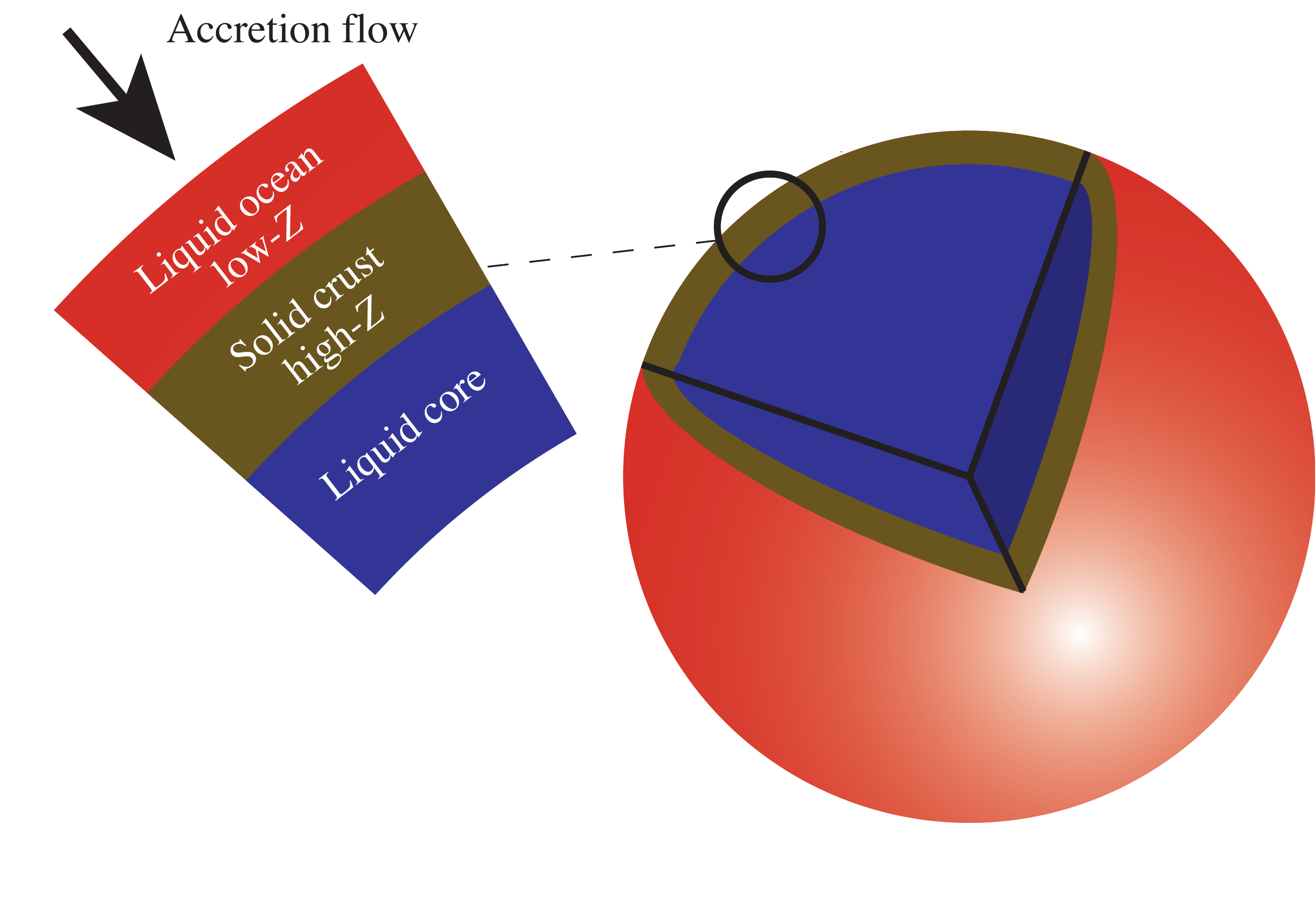}
\caption{Schematic diagram of the surface of an accreting neutron star \cite{phasesep}.  Chemical separation upon crystallization will take place at the boundary between the liquid ocean and the solid crust.  The ocean is enriched in low $Z$ elements.  } 
\label{Fig3}
\end{center}
\end{figure}

We now consider liquid-solid phase equilibria and chemical separation for more complicated compositions that may be present as accreting neutron stars form new crust.   With chemical separation, the liquid ocean is greatly enriched in low atomic number $Z$ elements, see Fig. \ref{Fig3}.    Carbon, if present, may be depleted in the crystal (crust) and enriched in the liquid ocean phase.   Also, chemical separation may change the thermal conductivity of the crust and its temperature profile.  Indeed some neutron stars are observed to produce energetic X-ray bursts known as superbursts.  These are thought to involve the unstable thermonuclear burning of carbon \cite{superbursts, superbursts2, superbursts3}.  However it is unclear how the initial carbon concentration is obtained and how the ignition temperature is reached. 

\begin{figure}[ht]
\begin{center}
\includegraphics[width=5in,angle=0,clip=true] {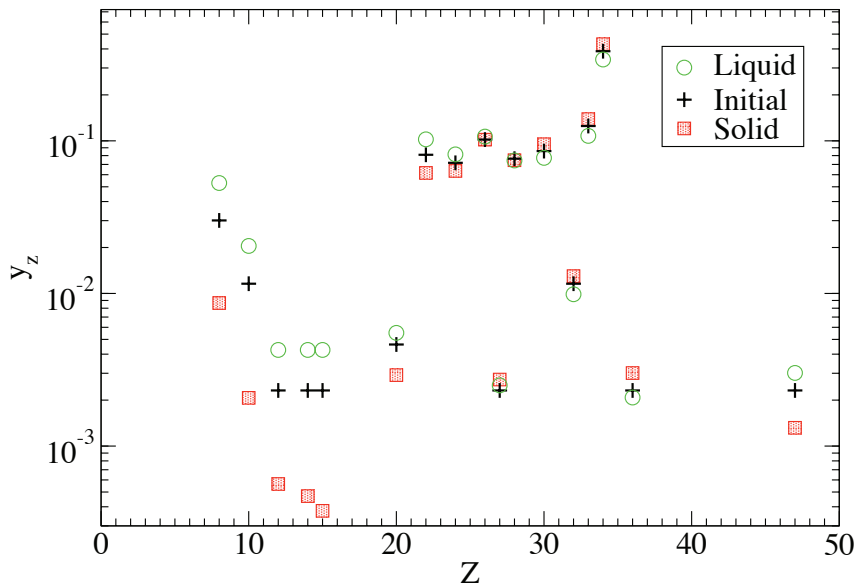}
\caption{Abundance (by number) of chemical elements versus atomic number $Z$ for rp process ash material on an accreting NS, from ref. \cite{phasesep}.  The plus symbols show the initial composition of the mixture.  The final compositions of the liquid phase, open green circles, and solid phase, filled red squares, are also shown.} 
\label{Fig4}
\end{center}
\end{figure}

Schatz et al. have calculated the rapid proton capture (rp) process of hydrogen burning on the surface of an accreting neutron star \cite{rpash}.  This produces a variety of nuclei up to atomic masses $A\approx 100$.  Gupta et al. \cite{gupta} then calculate how the composition of the rp process ash evolves because of electron capture and light particle reactions as the material is buried by further accretion.  Their final composition, at a density of $2.16\times 10^{11}$ g/cm$^3$ (near neutron drip at the bottom of the outer crust) has forty percent of the ions with atomic number $Z=34$, while an additional 10 \% have $Z=33$.  The remaining half of the ions have a range of lower $Z$ from $Z=8$ to 32.  Finally there is a small abundance of $Z=36$ and $Z=47$.  This complex composition is shown in Fig. \ref{Fig4}.
      
We have performed a two phase MD simulation using this complex 17 component rp-ash composition \cite{phasesep}.  The MD simulation is similar to the above carbon-oxygen simulation except that only 27648 ions were used.  The results are shown in Fig. \ref{Fig4}.  The liquid ocean is seen to be greatly enriched in low $Z$ elements such as oxygen compared to the solid crust.  Medin and Cumming have shown how this chemical seperation upon crystallization can lead to enrichment of the entire ocean \cite{medin_diffusion}.


\subsection{Diffusion}
\label{subsec.diffusion}
Diffusion of impurities, dislocations, or other defects can be very important for the structure of neutron star crust and could impact transport and mechanical properties.  Diffusion in coulomb plasma {\it liquids}  has been well studied \cite{Hanson75} and is important for sedimentation of impurities in white dwarf (WD) \cite{Ne_lars, Ne_lars2, neon_diffusion} and neutron stars (NS) \cite{NS_diffusion, peng}.  Here ions, with a larger than average mass to charge ratio, sink in a strong gravitational field.  This releases gravitational energy that can delay the cooling of metal rich WD \cite{WD_cooling}.  However, we are not aware of  previous numerical results for diffusion constants of coulomb {\it crystals} under Astrophysical conditions.  Often the diffusion constant is simply assumed to be zero.  This diffusion could be important for sedimentation in solid WD interiors, over long time scales, and for the structure of NS crusts.

Solid diffusion can depend dramatically on the form of the interaction between particles and may be very slow for hard-core systems.  For example, the binary Lennard Jones (LJ) system with a hard-core $\propto r^{-12}$ interaction forms a glass because of very slow diffusion \cite{LJglass}.  In contrast, the coulomb plasma with a soft $1/r$ core should have much faster diffusion.  Therefore the Coulomb crystal may provide an important model system where diffusion is fast enough to be more easily studied by molecular dynamics (MD) simulations.  

\begin{figure}[ht]
\begin{center}
\vskip.1in
\includegraphics[width=3.25in,angle=0,clip=true] {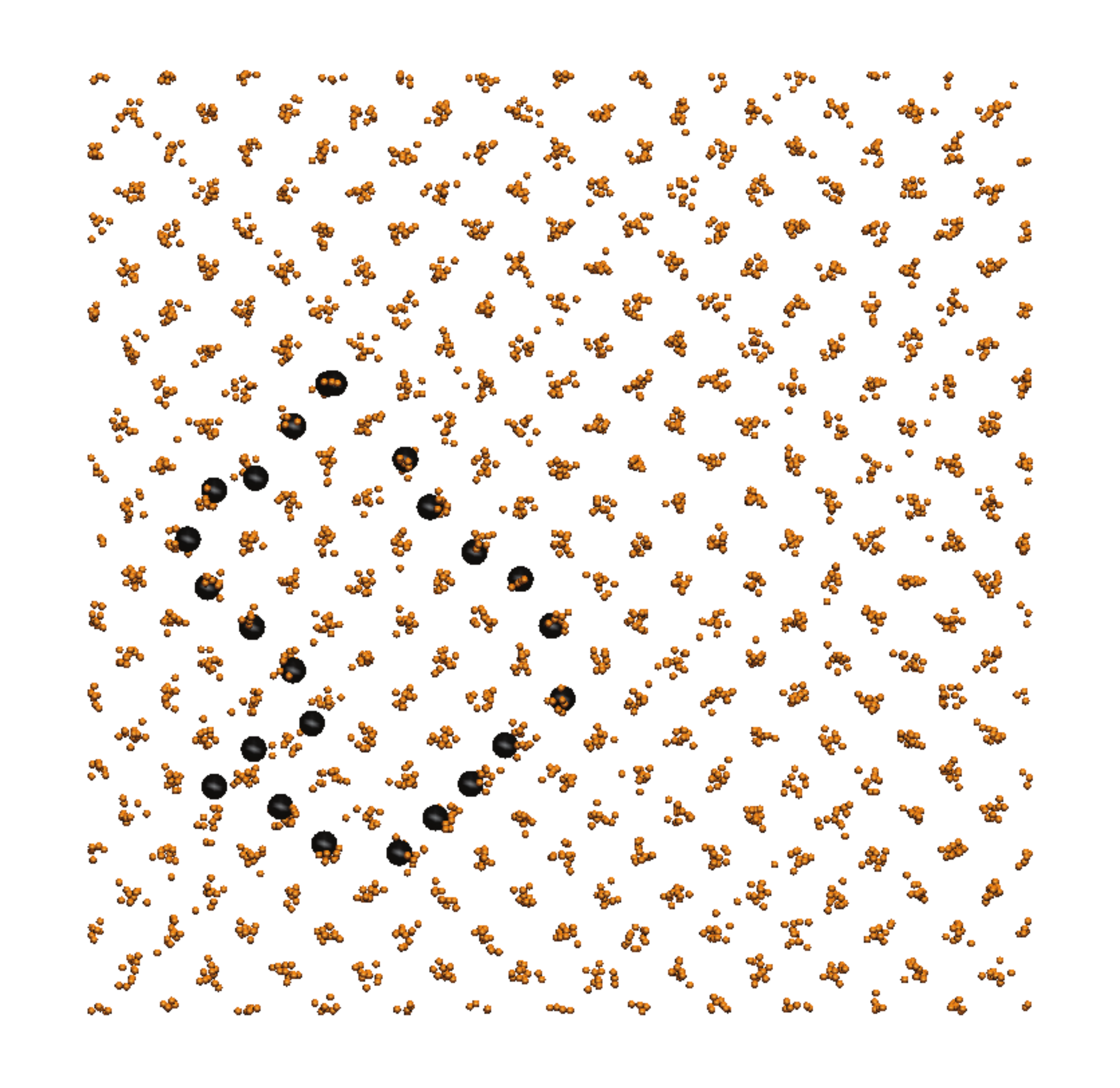}
\caption{Diffusion in a sample configuration of 3456 ions at $\Gamma=175$ \cite{soliddiffusion}.  Ions that have moved less than $1.34a$ in a time $t=236/\omega_p$ are small brown dots.  Ions that have moved more than $1.34a$ are shown as larger black disks and are seen to be in a ring configuration where ions ``hop'' to lattice sites vacated by other hopping ions.  This system started from a perfect bcc lattice.  Figure plotted with VMD \cite{VMD}. }
\label{Fig5}
\end{center}
\end{figure}

We have studied diffusion in solid one component plasmas \cite{soliddiffusion}.  For example, Fig. \ref{Fig5} shows ions that have diffused in a small 3456 ion simulation that started from a nearly perfect body centered cubic lattice.  The diffusing ions are seen to form a chain where ions hop to fill vacancies formed by other diffusing ions.   Neutron star crust is under great pressure and this suppresses vacancy formation.  Therefore ions move in ways to minimize vacancies.

\begin{figure}[ht]
\begin{center}
\vskip.1in
\includegraphics[width=3.25in,angle=0,clip=true] {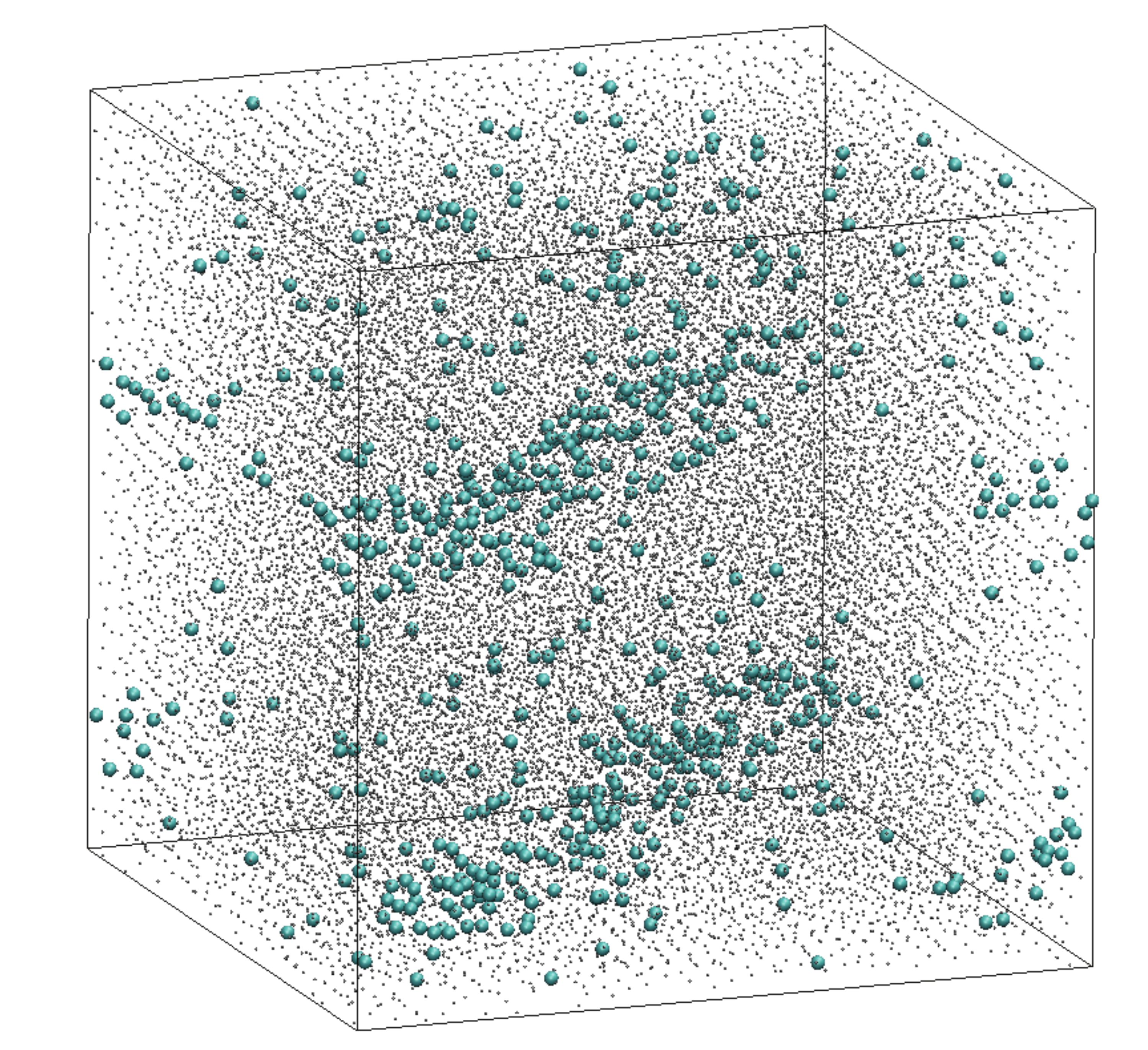}
\caption{Diffusion in a configuration of 27648 ions starting from imperfect crystal initial conditions \cite{soliddiffusion}.  Ions that move only a small distance are small gray points.  Ions that have moved over three lattice spacings, during the simulation time of $t=59000/\omega_p$, are shown as large blue spheres.  These are seen to be clustered at the grain boundaries.  The initial conditions included two micro-crystals of different orientation.  Figure plotted using VMD \cite{VMD}. }
\label{Fig6}
\end{center}
\end{figure}

Next in Fig. \ref{Fig6} we show diffusion in a 27648 ion system consisting of two micro-crystals of different orientations.  Ions are seen to diffuse quickly along the grain boundaries between micro-crystals.  Finally we considered diffusion in amorphous systems where a liquid configuration was quickly quenched from the melting temperature to a much lower temperature.  We find that diffusion in these systems is fast enough so that the systems can start to crystalize even over the relatively short time scales accessible to MD simulations.  This strongly suggests that coulomb solids in WD interiors and NS crust will be crystalline and not amorphous.
 
In summary, diffusion in Coulomb crystals is relatively fast because the ions have soft $1/r$ cores and can slide past one another.  As a result astrophysical solids in WD and NS are likely to be nearly perfect crystals with very few defects.  This suggests they should have a high thermal conductivity.  This is consistent with observations of rapid crust cooling of neutron stars following extended periods of accretion \cite{crustcooling1,crustcooling2,crustcooling3,crustcooling4}.  This rapid cooling implies a high crust thermal conductivity, that agrees with the conductivity of a regular crystal, and is larger than the conductivity expected for an amorphous solid.         


\subsection{Breaking Strain and Gravitational Waves}
\label{subsec.breaking}

The breaking strain (strength) of NS crust determines the maximum sized mountain that is possible on a NS before it collapses under the stars extreme gravity.  A large mountain, on a rapidly rotating NS, produces a time dependent mass quadrupole moment.  This efficiently radiates gravitational waves.   Albert Einstein, almost 100 years ago, predicted the oscillation of space and time known as gravitational waves (GW).  Within a few years, with the operation of Advanced LIGO \cite{advancedLIGO}, Advanced VIRGO \cite{advancedVIRGO} and other sensitive interferometers, we anticipate the historic detection of GW.  

The first GW that are detected will likely come from the merger of two neutron stars.  The rate of such mergers can be estimated from known binary systems \cite{LIGOrate}.   During a merger the GW signal has a so called chirp form where the frequency rises as the two neutron stars spiral closer together.  Deviations of this wave form from that expected for two point masses may allow one to deduce the equation of state of neutron rich matter and measure the radius of a neutron star $r_{NS}$ \cite{GWEOS}.   Alternatively one may be able to observe the frequency of oscillations of the hyper-massive neutron star just before it collapses to a black hole.  This frequency depends on the radius of the maximum mass neutron star \cite{Rmax}.  However, either approach may require high signal to noise data from relatively nearby mergers.  

Continuous GW signals can also be detected, see for example \cite{Collaboration:2009rfa}, in addition to burst signals such as those from neutron star mergers.  There are several very active ongoing and near future searches for continuous gravitational waves at LIGO, VIRGO and other detectors, see for example \cite{abbot}.  Often one searches at twice the frequency of known radio signals from pulsars because of the quadrupole nature of GW.  No signal has yet been detected.  However, sensitive upper limits have been set.  These limits constrain the shape of neutron stars.  In some cases the star's elipticity $\epsilon$, which is that fractional difference in moments of inertia $\epsilon=(I_1-I_2)/I_3$ is observed to be less than a part per million or even smaller.  Here $I_1$, $I_2$, and $I_3$ are the principle moments of inertia.    

In general, the amplitude of any continuous signal is much weaker than a burst signal.  However, one can gain sensitivity to a continuous signal by coherently (or semi-coherently) integrating over a large observation time, see for example \cite{semicoherent}.  Note that searches for continuous GW can be very computationally intensive because one must search over an extremely large space of parameters that may include the source frequency, how that frequency changes with time, the source location on the sky, etc.  The Einstein at home distributed computing project uses spare cycles on the computers of a large number of volunteers to search for continuous GW \cite{Einstein@home}. 
           
Strong GW sources often involve large accelerations of large amounts of neutron rich matter.  Indeed the requirements for a strong source of continuous GW, at LIGO frequencies, places extraordinary demands on neutron rich matter.   Generating GW sounds easy.  Place a mass on a stick and shake vigorously.  However to have a detectable source, one may need not only a large mass, but also a very strong stick.   The stick is needed to help produce large accelerations.  

An asymmetric mass on a rapidly rotating neutron star produces a time dependent mass quadrupole moment that radiates gravitational waves.  However, one needs a way (strong stick) to hold the mass up.  Magnetic fields can support mountains, see for example \cite{magneticmountains}.  However, it may require large internal magnetic fields.  Furthermore, if a star also has a large external dipole field, electromagnetic radiation may rapidly spin the star down and reduce the GW radiation.  

\begin{figure}[h]
\center\includegraphics[width=4.25in]{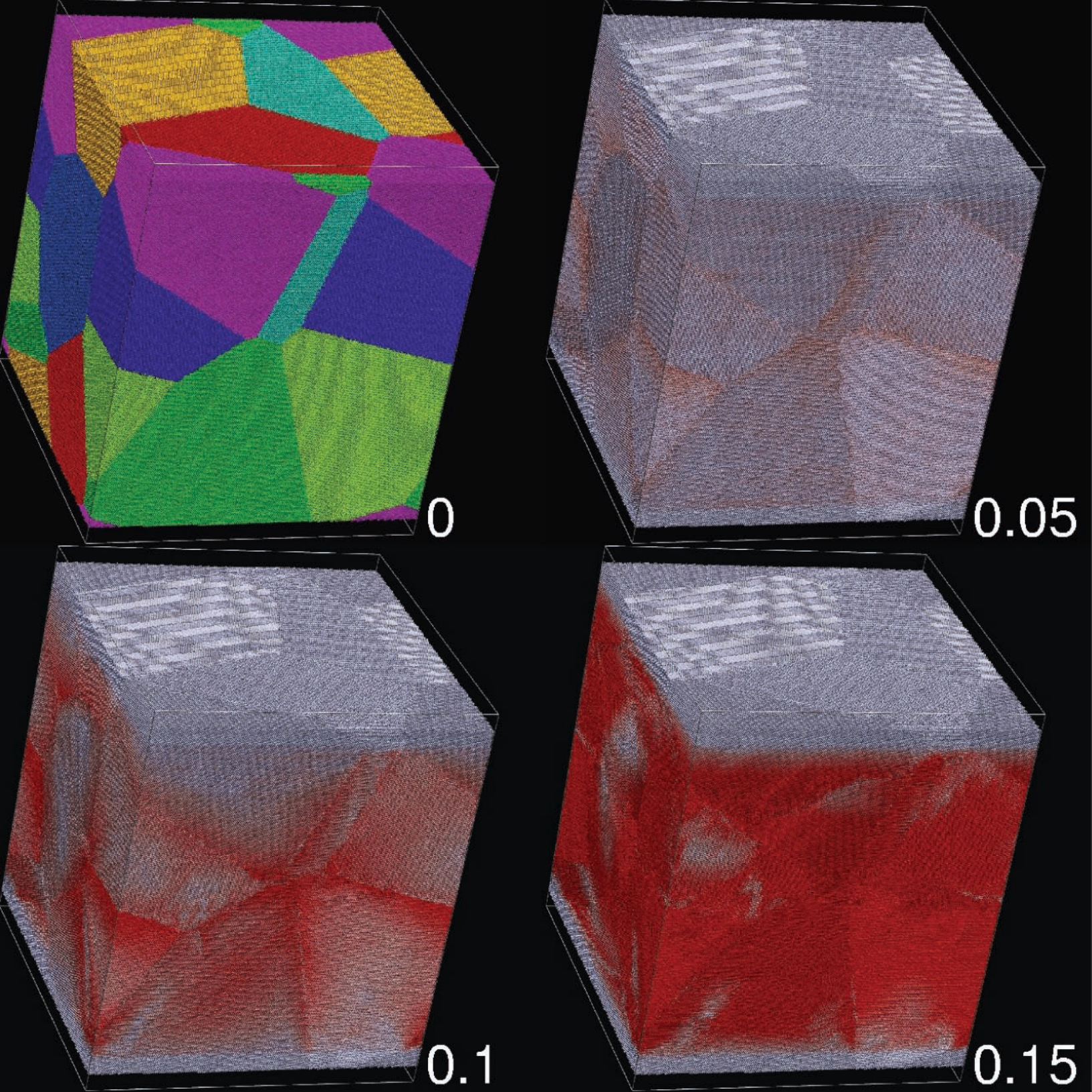}
\caption{\label{Fig7}  Breaking strain (strength) of neutron star crust from MD simulations \cite{crustbreaking}.  The shear deformation of a 12 million ion polycrystaline sample is shown.  The panels are labeled by the strain, which is the fractional deformation.   The colors in the upper left panel, at zero strain, show the original eight microcrystals of different orientations.  The red color in the other three panels indicates distortion of the body centered cubic lattice.}
\end{figure}

Alternatively, mountains can be supported by the solid neutron star crust.
Recently we performed large scale MD simulations of the strength of neutron star crust \cite{crustbreaking,chugunov}.   A strong crust can support large deformations or ``mountains'' on neutron stars, see also \cite{lowmassNS}, that will radiate strong GW.   How large can a neutron star mountain be before it collapses under the extreme gravity?  This depends on the strength of the crust.  We performed large scale MD simulations of crust breaking, where a sample was strained by moving top and bottom layers of frozen ions in opposite directions \cite{crustbreaking}.  These simulations involve up to 12 million ions and explore the effects of defects, impurities, and grain boundaries on the breaking stress.  For example, in Fig. \ref{Fig7} we show a polycrystalline sample involving 12 million ions.  In the upper right panel the initial system is shown, with the different colors indicating the eight original microcrystals that make up the sample.  The other panels are labeled with the strain, i.e. fractional deformation, of the system.  The red color indicates distortion of the body centered cubic crystal lattice.  The system starts to break along grain boundaries.  However the large pressure holds the microcrystals together and the system does not fail until large regions are deformed.

We find that neutron star crust is very strong because the high pressure prevents the formation of voids or fractures and because the long range coulomb interactions insure many redundant ``bonds'' between planes of ions.  Neutron star crust is the strongest material known, according to our simulations.  The breaking stress is 10 billion times larger than that for steel.  This is very promising for GW searches because it shows that large mountains are possible, and these could produce detectable signals.  

\subsection{Shear viscosity of Nuclear Pasta and r-modes}
\label{subsec.viscosity}

Continuous GW can also be produced by r-mode oscillations of a rotating neutron star \cite{r-modereview}.  Consider a surface wave on a rapidly rotating neutron star that is moving slowly in a direction opposite to the stars rotation.  This wave, in the laboratory frame, will appear to be moving in the direction of the star's rotation.  The back reaction force on the wave from GW radiation will always act to slow the wave in the laboratory frame.  However, in this case slowing in the lab frame will actually speed up the wave in the rotating star's frame and increase its amplitude.  Thus the wave can be unstable with respect to gravitational wave radiation.   The r-modes are collective oscillations on rotating neutron stars that can also be unstable to GW radiation \cite{r-modereview}.  If an r-mode is unstable the amplitude will grow large and rotational kinetic energy can be radiated away as GW.  This will slow the rotation rate.  Note, that the physics of large amplitude r-mode oscillations can be complicated, see for example \cite{largermode,largermode2}.

The stability of r-modes depends on the amount of dissipation from, for example, the bulk and shear viscosities of neutron rich matter.  If dissipation is large then the amplitude of the r-modes will stay small and rapid rotation of a neutron star is possible.  Alternatively, if dissipation is small then the r-modes may be unstable and GW radiation from the modes may limit the rotation rate of a star.  Unfortunately, the stability of r-modes has proved to be a complex subject that may be sensitive to subtle dissipation properties of neutron rich matter, be it in a nucleon phase \cite{shearviscosity1,shearviscosity2} or in more exotic quark and gluon phases \cite{kaonbulk, quarkbulk, quarkviscosity}.

\begin{figure}[h]
\center\includegraphics[width=4in]{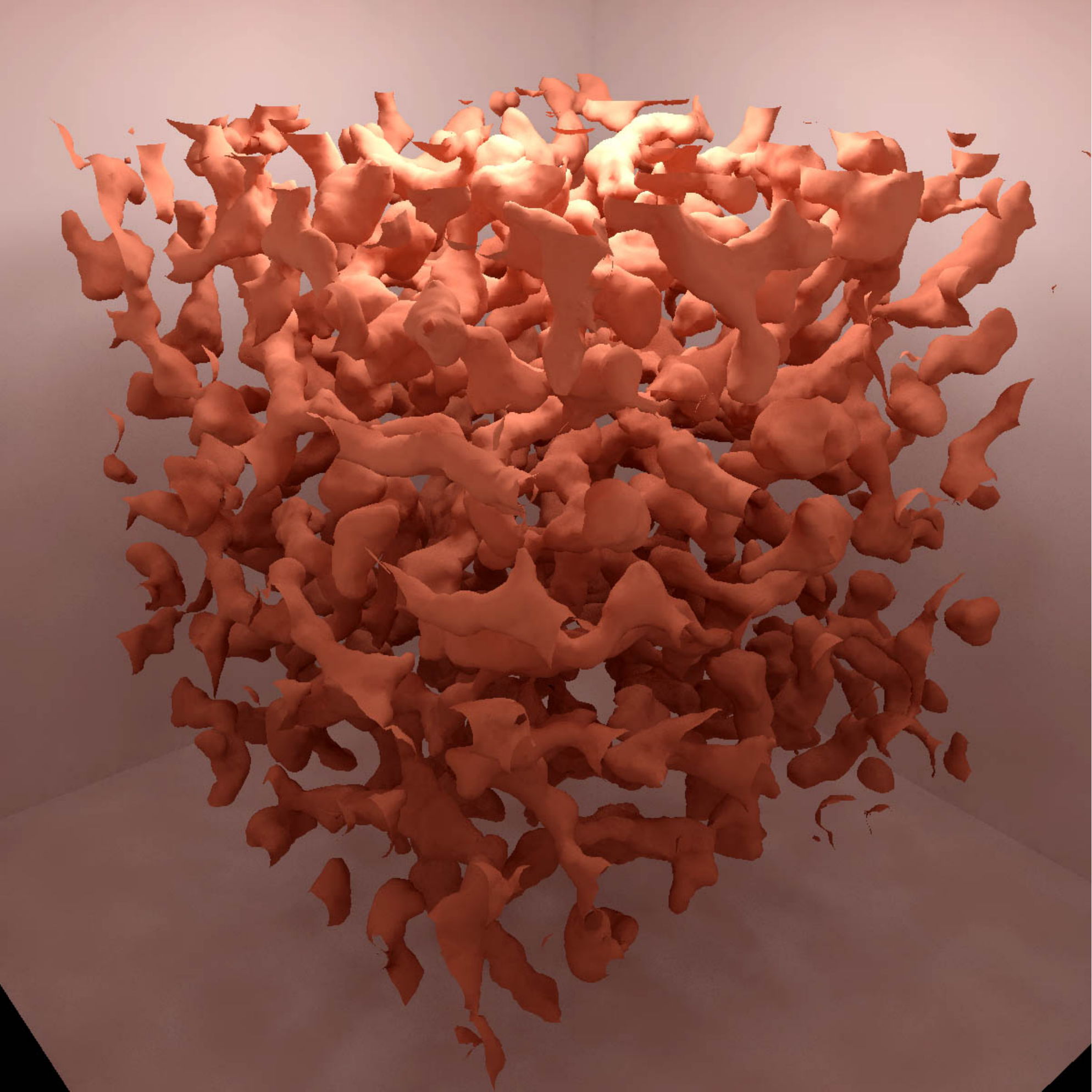}
\caption{\label{Fig8}Surfaces of proton density for a pasta configuration of neutron rich matter at a baryon density of 0.05 fm$^{-3}$.  This is from a semiclassical molecular dynamics simulation with 100,000 nucleons \cite{pasta2}.} 
\end{figure}

We give one example of a possible source of dissipation for the r-modes.  The shear viscosity of conventional complex fluids, with large non-spherical molecules, can be orders of magnitude larger than that for normal fluids.  This suggests that the shear viscosity of nuclear pasta, with complex non-spherical shapes such as the long bent rods shown in Fig. \ref{Fig8}, could be large.  Complex shapes arise because of coulomb frustration \cite{pasta}.  Here one can not fully satisfy both short range nuclear attraction, that correlates nucleons, and long range Coulomb repulsion, that anti-correlates nucleons.  Pasta is expected at the base of the crust in a neutron star and can involve a variety of complex shapes such as flat plates (``lasagna'') in addition to long rods (``spaghetti'').

In ref. \cite{pastaviscosity}, we have calculated the shear viscosity of nuclear pasta using large scale molecular dynamics simulations.  The shear viscosity is dominated by momentum carried by electrons and although the electron mean free path is determined by electron-pasta scattering, we find no dramatic differences from a conventional phase with spherical nuclei.  Therefore, we find that the shear viscosity of nuclear pasta is not very different from that for more conventional matter with nearly spherical nuclei.  However there could be other sources of dissipation.  We do not yet have a complete understanding of when the r-modes may be stable and when they are unstable.  

\subsection{Dynamical Response of Nuclear Pasta}
\label{subsec.pastaosc}
The low energy excitation modes of nuclear pasta are important for the heat capacity, and a variety of transport properties, of matter deep in the inner crust.  There have been some RPA calculations of excitation modes, see for example \cite{pastaRPA, pastaRPA2}.  Unfortunately many of these works use spherical Wigner Seitz cell boundary conditions that could dramatically impact low energy excitations.  Semiclassical MD simulations allow one to calculate the dynamical response function, see below, and the excitation spectrum without any assumptions about unit cell geometries, \cite{pastaresponse}.

As an example, we consider the neutrino opacity in core collapse supernovae.   Neutrinos may scatter coherently from the pasta shapes because the shapes have sizes comparable to the neutrino wavelength \cite{pastascattering}.   The differential cross section for neutrino scattering may be written as follows~\cite{pastaresponse}:
\begin{equation}
 \frac{d\sigma}{d\Omega dE}=\frac{G_F^2E_\nu^2}{4\pi^2} 
 \Bigl[c_a^2(3-x )S_A(q,\omega) + c_v^2(1+x)S_{V}(q,\omega)\Bigr],
 \label{sigma}
\end{equation}
where $G_F$ is the Fermi constant, $E_\nu$ is the neutrino energy,
$x=\cos\theta$ (with $\theta$ the scattering angle), and the weak
vector charge of a nucleon is $c_v\!=\!-1/2$ for a neutron and
$c_v\!\approx\!0$ for a proton. Further, the dynamic response functions $S_V(q,\omega)$, $S_A(q,\omega)$ are probed at a momentum transfer $q$ and at an energy transfer $\omega$. The axial term involving $c_a\!=\!\pm 1.26/2$ and the
dynamical spin response $S_A(q,w)$ will be discussed in a later work.
Here we focus exclusively on the vector (density) response
$S(q,\omega)\!\equiv\!S_{V}(q,\omega)$ that should be greatly enhanced
by coherent effects.

The dynamical response of the system to a density perturbation
is given by
\begin{equation}
  S(q,\omega)=\frac{1}{\pi} \int_0^{T_{\rm max}} 
               S(q,t) \cos(\omega t)\, dt \;.
\label{sqw}
\end{equation}
Here $S(q,t)$ represents the ensemble average of the density-density 
correlation function that is computed as the following time average:  
\begin{equation}
  S(q,t)=\frac{1}{N}\frac{1}{T_{\rm ave}}\int_0^{T_{\rm ave}} 
         \rho({\bf q},t+s)\rho(-{\bf q},s) ds \;.
 \label{sqt}
\end{equation}
In the above expression $N$ is the number of neutrons in the system 
and appropriate choices for $T_{\rm max}$ and $T_{\rm ave}$ depend on the MD simulation \cite{pastaresponse}.  Finally, the one-body neutron density is given by $\rho({\bf q},t)=\sum_{i=1}^N 
 \exp[{i{\bf q}\cdot {\bf r}_i(t)}]$
with ${\bf r}_i(t)$ the position of the $i_{}$th neutron at time 
$t$. Note that because $c_v\!\approx\!0$ for protons, the sum over
$i$ runs only over neutrons. Further, the static 
structure factor computed in Ref.~\cite{pasta2} is easily recovered 
from $S(q)\equiv S(q,t\!=\!0)=\int_0^\infty S(q,\omega) d\omega$. 

\vspace{0.02in}
\begin{figure}[ht]
\begin{center}
\includegraphics[width=5in,angle=0,clip=false]{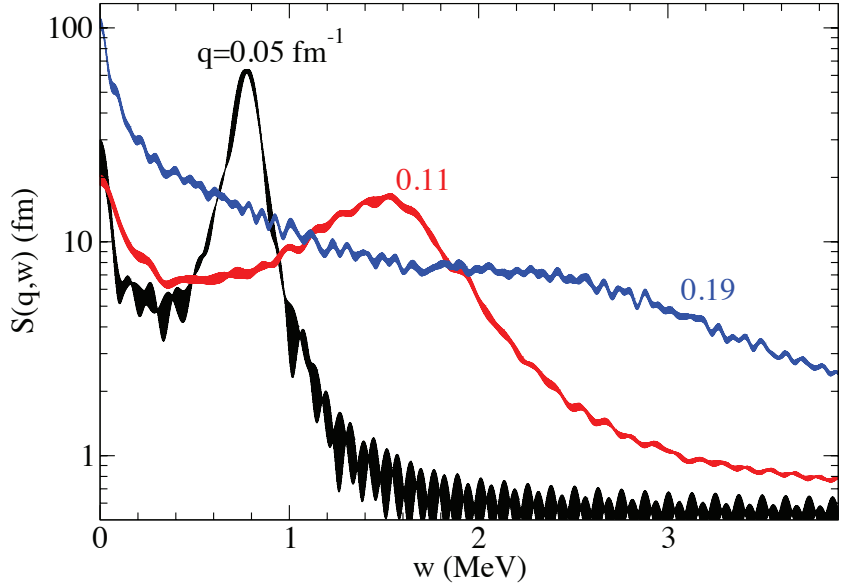}
\caption{(Color online) The dynamical response function  
          $S(q,\omega)$ of nuclear pasta versus excitation energy $\omega=$w 
          at a density of $\rho\!=\!0.05$ fm$^{-3}$ and 
	  momentum transfers of $q\!=\!0.05$, 0.11, and 
	  0.19 fm$^{-1}$.  Figure from ref. \cite{pastaresponse}.}
\label{Fig9}
\end{center}
\end{figure}

In Fig. \ref{Fig9} we show $S(q,\omega)$ for nuclear pasta at a temperature of 1 MeV \cite{pastaresponse}.  The $q=0.05$ fm$^{-1}$ response shows a large peak near $\omega=0.8$ MeV that is due to plasma oscillations of the charged pieces of pasta.  The plasma frequency $\omega_p$ depends on the charge and mass of the pasta, see for example Eq. \ref{omega}, however $\omega_p$ appears to be relatively insensitive to the pasta shape.  The peak in Fig. \ref{Fig9}  shifts to higher excitation energies and becomes broader as $q$ increases.  The frequency goes up because electron screening is less effective at higher $q$, while the larger width may describe an increased coupling between Coulomb and nuclear excitations.

Nuclear pasta arrises because of competition between long range Coulomb repulsion and short ranged nuclear attraction.  Indeed at these densities there is some overlap between typical excitation energy scales of the Coulomb lattice and typical nuclear excitation energies.  This is reflected in the plasma oscillation peak.  This peak could be important for the heat capacity and for transport properties.  

\section{Summary, Open Questions, and Challenges for the Future}
\label{sec.conclusions}
Neutron star crust consists of neutron rich ions, electrons, and possibly a neutron gas.  The electronic structure is simple, a degenerate relativistic gas, because of the very high Fermi energy.  As a result, molecular dynamics simulations provide an essentially exact description and can predict many crust properties.  We find that diffusion is relatively fast in the crust because of the soft $1/r$ Coulomb cores between ions.  This strongly suggests that amorphous structures will have plenty of time to crystallize, and many crystal defects will diffuse away.  Therefore we expect neutron star crust to be made of nearly perfect crystals with high electrical and thermal conductivities.  Furthermore, we find that these crystals are very strong because of the many long range Coulomb interactions.

There are important open questions concerning the composition of the crust.  For an isolated star the composition depends on nuclear interactions including the symmetry energy.  While for accreting stars the composition also depends on the rates of a variety of nuclear reactions including pycnonuclear (density driven) fusion \cite{pycno}.       

A strong crust can support large mountains.  These, on a rapidly rotating star, can generate detectable gravitational waves.  An important challenge is to understand mountain building mechanisms and determine what sized mountains are, in fact, present.  This is important for many ongoing and near future searches for continuous gravitational waves.

There are many open questions and challenges associated with nuclear pasta.  Perhaps the most basic is "how to smell the pasta?".  Which neutron star observables are sensitive to pasta, and which ones can demonstrate the presence of complex non-spherical pasta shapes.  What is the shear modulus and breaking strain of pasta?  What dissipation mechanisms are associated with oscillations of pasta?

Finally, how do neutron stars spin so fast and what limits there rotational periods?  We do not understand the stability of r-modes.  The amplitude of r-mode oscillations is large in the crustal regions.  Perhaps there are new sources of dissipation in the crust, such as bulk viscosity from an electron capture layer,  that stabilize r-modes and allow rapid rotation.

\section*{Acknowledgments}

This work was supported in part by DOE grant DE-FG02-87ER40365 and by the National Science Foundation, TeraGrid grant TG-AST100014.

\label{lastpage-01}

\end{document}